\documentclass[twocolumn,twoside,slac_two]{revtex4}

\usepackage{graphicx}
\usepackage{fancyhdr}
\usepackage{graphics}
\usepackage{epstopdf}
\usepackage{textpos}
\def\pt{\ensuremath{p_{\mathrm{T}}}}

\def\MET{\ensuremath{E_{\mathrm{T}}^{\mathrm{miss}}}} 
\def\met{\ensuremath{E_{\mathrm{T}}^{\mathrm{miss}}}}
\newcommand{\Ztt}{\ensuremath{Z \to \tau\tau}}
\newcommand{\Mll}{\ensuremath{m_{\ell\ell}}}
\newcommand{\mz}{\ensuremath{m_{Z}}}
\def\antibar#1{\ensuremath{#1\bar{#1}}}
\def\ttbar{\antibar{t}}

\begin{document}

%%%%%%%%%%%%%%%%%%%%%% WRITE THE TITLE HERE %%%%%%%%%%%%%%%%%%%
\title{\centering Measurement of Single-top Quark Production with ATLAS Data}
%%%%%%%%%%%%%%%%%%%%%% WRITE THE AUTHOR HERE %%%%%%%%%%%%%%%%%
\author{
\centering
\begin{center}
J. L. Holzbauer on behalf of ATLAS
\end{center}}
\affiliation{\centering Michigan State University, MI, 48864, USA}
%%%%%%%%%%%%%%%%%%%%%% WRITE THE ABSTRACT HERE %%%%%%%%%%%%%%%%
\begin{abstract}
Single-top production processes have been studied using 0.7 $\mathrm{fb^{-1}}$ of data from 7 TeV center-of-mass energy proton-proton collisions collected with the ATLAS detector at the LHC. Single-top is electroweak top production and the standard model includes three production modes. Each contains a Wtb vertex, allowing the possibility of a direct measurement of the CKM matrix element $|V_{tb}|$. Single-top could also be sensitive to new physics, such as flavor changing neutral currents or heavy $W'$ bosons. Using cut-based selections, a limit of $\sigma_{Wt} < 39.1$ pb is set for dilepton $Wt$ production and $\sigma_{s} < 26.5$ pb for $s$-channel production. For the $t$-channel measurement, both cut-based and neural network analyses are performed and the cross-section is measured to be $90^{+32}_{-22}$ pb, where $65^{+28}_{-19}$ pb is expected according to standard model. 
\end{abstract}

\maketitle
\thispagestyle{fancy}


\section{Introduction}
Single-top is electroweak top-quark production through three different standard model production modes: $t$-channel, $Wt$ (associated production) and $s$-channel. The $t$-channel has the largest expected cross-section for 7 TeV proton-proton collisions ($64.6^{+3.3}_{-2.6}$ pb)~\cite{Kidonakis:2011wy} followed by $Wt$ ($15.7\pm1.4$ pb)~\cite{Kidonakis:2010ux}, and $s$-channel ($4.6\pm0.3$ pb)~\cite{Kidonakis:2010tc}. The $t$-channel and $Wt$ cross-sections in particular are at least 30 times larger at the LHC than at the Tevatron, while the $s$-channel cross-section is about 4 times as large~\cite{Kidonakis:2006tev}. The larger cross-sections will allow precision measurements of the cross-sections, tests of the Wtb coupling, direct measurements of the CKM matrix element $|V_{tb}|$~\cite{CKM1,CKM2}, and searches for new physics such as flavor changing neutral currents and heavy bosons~\cite{Tait:2000sh}.  

Single-top production was first observed at the Tevatron in 2009 by the D0~\cite{Abazov:2009ii} and CDF~\cite{Aaltonen:2009jj} collaboration, and the $t$-channel production mode of single-top alone was first observed by D0~\cite{Abazov:2011rz}. Here, results from ATLAS~\cite{ATLASdet}, a detector located at the LHC, are discussed. Of the data collected, 0.7 $\mathrm{fb^{-1}}$ is used to measure the $t$-channel~\cite{ATLAS-CONF-2011-101} cross-section, set a $Wt$~\cite{ATLAS-CONF-2011-104} limit, and set the first ATLAS $s$-channel~\cite{ATLAS-CONF-2011-118} limit.

\section{\textit{t}-channel Analysis}\label{sec:tchan}
The $t$-channel analysis is performed using two different techniques, a cut-based selection and a neural network approach. In both cases the analyses begin by reducing the large multijet and $W$+jets backgrounds via a common selection which forms a sample enriched with events containing a $t$-channel final state. Events with at least 2-jets with transverse momentum of $\pt > 25~\mathrm{GeV}$ and $|\eta| < 4.5$ are selected, where exactly 1 jet must be b-tagged using a secondary vertex b-tagging algorithm with a 50\% b-tagging efficiency and a rejection of about 270 for light-quarks. The analyses are performed using 2-jet or 3-jet multiplicity bins. Events are selected which have exactly 1 isolated, central ($|\eta| < 2.5$) muon or electron with $\pt > 25~\mathrm{GeV}$ which is associated with a corresponding lepton trigger. The events are also required to have missing transverse energy, $\met > 25~\mathrm{GeV}$, to account for the undetected neutrino from the $W$ decay. Finally, $\met + M_{\mathrm{T}}(W) > 60~\mathrm{GeV}$ is required to further reduce multijet contributions, where $M_{\mathrm{T}}(W)$ is the reconstructed transverse mass of the $W$ determined from the lepton and \met. After all of these selections, the signal yield divided by the background yield (S/B) is approximately 0.1.

The background normalization uses theoretical cross-sections for most processes (\ttbar, $Wt$, $s$-channel, diboson, and $Z$+jets backgrounds) except the large $W$+jets and multijet backgrounds. In the case of multijets, the kinematic distribution shapes are determined using a ``jet-electron'' method, where the typical electron trigger is replaced by a jet trigger, and the normalization is determined via a binned maximum likelihood fit to the \met~distribution in data~\cite{Aad:2010ey_sgtop}. This method is also used for the muon event selection. The $W$+jets heavy flavor fractions and normalization are determined using three control regions in the cut-based analysis and via a fit of the neural network output distribution in that analysis.

\begin{figure}
\centering
\includegraphics[width=75mm]{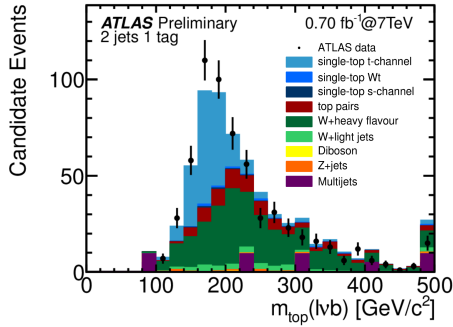}
\caption{Reconstructed top mass distribution for 2-jet events after all $t$-channel cut-based analysis selections except the reconstructed top mass selection~\cite{ATLAS-CONF-2011-101}. The $t$-channel signal contribution is normalized to the measured combined cut-based analysis $t$-channel cross section.}
\label{fig:topmassafterpresel}
\end{figure}

The cut-based analysis applies four selections in addition to the initial selection: $|\eta({\rm ljet})| > 2.0$, $H_{\mathrm{T}}>210~\mathrm{GeV}$, $150~\mathrm{GeV} < M_{l\nu b} < 190~\mathrm{GeV}$ and $|\Delta\eta({\rm bjet}, {\rm ljet})| > 1.0$. Here, ljet refers to the untagged jet with the largest \pt~value and bjet refers to the b-tagged jet. Additionally, $H_{\mathrm{T}}$ is sum of the transverse momentum of the leading two jets in \pt, the lepton \pt, and \met~while $M_{l\nu b}$ is the reconstructed top quark mass. Four channels are used in this analysis, containing 2 or 3-jets and positively or negatively charged leptons. This channel division makes use of the expected lepton charge asymmetry in p-p collisions due to the excess of valence u-quarks versus d-quarks. Such a division would be ineffective at a p-$\mathrm{\bar{p}}$ collider, where no lepton asymmetry is expected. After the cut-based analysis selections, the best channel (2-jets with positive leptons) has a S/B value of 1.2. The distribution of the reconstructed top mass for 2-jets (combined lepton charge channels) is shown in Figure~\ref{fig:topmassafterpresel} and the signal excess can be clearly seen.

A neural network analysis is also performed using the NeuroBayes~\cite{feindt-2004,Feindt:2006pm} program with 13 variables, including the variables used in the cut-based analysis and lepton charge, used in its channel division. The other 8 variables include the invariant mass of the two jets, $M_{\mathrm{T}}(W)$, the lepton $\eta$ and \pt, \met, the \pt~of the untagged jet, the mass of the b-tagged jet, and $|\Delta\eta|$ between the b-tagged jet and reconstructed $W$ boson. The variables are combined, including correlations, into one discriminant. Unlike the cut-based selection, no additional selections are applied after initial selection and all events contain exactly 2-jets.

The $t$-channel cross-sections are extracted using a profile likelihood technique for the cut-based analysis (CB) and a maximum likelihood fit for the neural network analysis (NN). The inclusive cross-sections obtained from the various methods are $\sigma_{t}= 102^{+40}_{-30}$ pb (CB 2-jets), $\sigma_{t}= 105^{+37}_{-31}$ pb (NN), and $\sigma_{t}=50^{+34}_{-27}$ pb (CB 3-jets), all of which are consistent with each other within at least two standard deviations.  

The final result is constructed as the combination of the two and three jet channels from the cut-based analysis, $\sigma_{t}= 90^{+9}_{-9}(\mathrm{stat})\,^{+31}_{-20}(\mathrm{syst})= 90^{+32}_{-22}$ pb, compared to $\sigma_{t}^\mathrm{exp}=65^{+28}_{-19}$ pb expected. The measured cross-section is consistent with the predicted standard model $t$-channel cross-section within about 1.1 standard deviations. The uncertainty of this measurement is dominated by systematic uncertainties, with b-tagging, jet energy scale, and initial/final state radiation being particularly large contributors to the cross-section uncertainty.

\section{\textit{Wt} Analysis}\label{sec:wtchan}
The $Wt$ analysis considers the dilepton final state only, using three different channels: electron-electron, electron-muon, and muon-muon. For this analysis, at least one jet is required with $\pt > 30~\mathrm{GeV}$, as well as exactly two isolated, triggered, central muons or electrons with $\pt > 25~\mathrm{GeV}$ and corresponding $\met > 50~\mathrm{GeV}$. Additionally, the sum of the angular differences in $\phi$ between the leptons and \met~direction is used to reduce the \Ztt~contamination, $\Delta\phi(l_1,\MET)+\Delta\phi(l_2,\MET) > 2.5$, and the reconstructed dilepton invariant mass is used to remove events coming from $Z$ boson decays, $|\Mll - \mz| > 10~\mathrm{GeV}$, for events where both leptons are electrons or both are muons. Here, $\Mll$ refers to the reconstructed mass of the two leptons and $\mz$ refers to the $Z$ boson mass.

\begin{figure}[h]
\centering
\includegraphics[width=65mm]{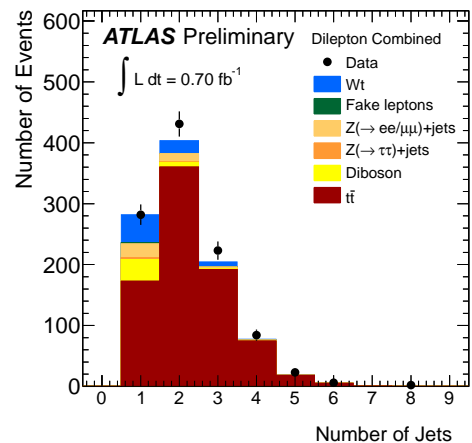}
\caption{Number of jets after initial $Wt$ event selection~\cite{ATLAS-CONF-2011-104}.  The 1-jet bin is used for the final selection, while the 2 or more jets region is used for the \ttbar~background normalization.}
\label{fig:WtSel}
\end{figure}

The background cross-section normalization is determined via data-based techniques with the exception of the diboson background, for which the theoretical cross-section is used. A matrix method is used to estimate the fake lepton (multijet and $W$+jets) background. The \Ztt~is estimated using a \Ztt~dominated region orthogonal to the signal region, $\Delta\phi(l_1,\MET)+\Delta\phi(l_2,\MET) < 2.5$, while orthogonal data samples formed using two uncorrelated variables, \met~and \Mll, are used to determine the Drell-Yan background. The largest background, \ttbar, is determined using a \ttbar~dominated region orthogonal to the signal region defined as events with at least 2-jets (see Figure~\ref{fig:WtSel}). The non-\ttbar~events in this region are subtracted from the data and the result is compared to the expected value to determine a scale factor. This result is then propagated into the signal region. Correlated effects of systematic uncertainties affecting both \ttbar~and $Wt$ have been taken into account.

The final $Wt$ selection requires exactly 1 jet (see Figure~\ref{fig:WtSel}). The cross-section and limit are extracted using a profile likelihood technique. The resulting cross-section is $\sigma_{Wt} = 14.4 ^{+5.3}_{-5.1} \mathrm{(stat)} ^{+9.7}_{-9.4} \mathrm{(syst)}$ pb. The 95\% confidence level (CL) upper limit on $Wt$ production is $\sigma_{Wt} < 39.1$ pb observed, $\sigma_{Wt} < 40.6 $ pb expected.  The total uncertainty is dominated by data statistics, jet energy scale, jet energy resolution, and jet reconstruction efficiency uncertainties.

\section{\textit{s}-channel Analysis}
The $s$-channel analysis event selection is similar to the $t$-channel cut-based analysis selection (see Section~\ref{sec:tchan}) with two differences: 2 central ($|\eta| < 2.5$) jets are selected and at least 1 b-tagged jet is required. The backgrounds are also estimated using the same techniques, but selecting events with at least 1 b-tagged jet. The cut-based selection includes a requirement of exactly 2 b-tagged jets, as well as $30~\mathrm{GeV} < M_{\mathrm{top,bjet2}} < 247~\mathrm{GeV}$, $\pt(\mathrm{bjet1}, \mathrm{bjet2})<189~\mathrm{GeV}$, $M_{\mathrm{T}}(W) < 111~\mathrm{GeV}$, $0.43 < \Delta R(\mathrm{bjet1}, \mathrm{lepton}) > 3.6$, $123~\mathrm{GeV} < M_{\mathrm{top, bjet1}} < 788~\mathrm{GeV}$, and $0.74 < \Delta R(\mathrm{bjet1}, \mathrm{bjet2}) > 4.68$. Here, $M_{\mathrm{top},\mathrm{bjet1}}$ is the reconstructed top quark mass using the leading jet in pt (bjet2 is the other jet). After these selections, S/$\mathrm{\sqrt{B}}$ is 0.98, an improvement from 0.26 using the initial selection only.

The $s$-channel cross-section and limit are extracted using a profile likelihood technique. The 95\% CL upper limit on the production of $s$-channel events is $\sigma_{s} < 26.5$ pb observed, where $\sigma_{s} < 20.5$ pb is expected. This is about 5 times the standard model cross-section. More data statistics will improve this result as data statistics is the dominant uncertainty.

\section{Conclusion}
ATLAS single-top production results are estimated using 0.7 $\mathrm{fb^{-1}}$ of data from 7 TeV proton-proton collisions. The $t$-channel cross-section is measured to be $90^{+32}_{-22}$ pb. The $Wt$ analysis determines a cross-section of $\sigma_{Wt} = 14.4 ^{+5.3}_{-5.1} \mathrm{(stat)} ^{+9.7}_{-9.4} \mathrm{(syst)}$ pb, corresponding to a 95\% CL limit of $\sigma_{Wt} < 39.1$ pb. Finally, a 95\% CL limit of $\sigma_{s} < 26.5$ pb is set for the single-top $s$-channel production cross-section.

\bigskip % extra skip inserted
%% Create the reference section using BibTeX:

\end{document}